\documentclass[11pt]{article}
\usepackage{latexsym,epic,eepic}
\usepackage{amssymb}
\usepackage{geometry}
\usepackage{amsmath}
\usepackage{amsfonts}
\usepackage{graphicx}

\begin{document}
\newcommand{\2}{\vspace{0.2 cm}}
\newcommand{\dist}{{\rm dist}}
\newcommand{\diam}{{\rm diam}}
\newcommand{\rad}{{\rm rad}}
\newcommand{\dom}{\mbox{$\rightarrow$}}
\newcommand{\ndom}{\mbox{$\not\rightarrow$}}
\newcommand{\sdom}{\mbox{$\Rightarrow$}}
\newcommand{\nsdom}{\mbox{$\not\Rightarrow$}}
\newcommand{\qed}{\hfill$\diamond$}
\newcommand{\pf}{{\bf Proof: }}
\newtheorem{theorem}{Theorem}[section]
\newcommand{\ra}{\rangle}
\newcommand{\la}{\langle}
\newtheorem{lemma}[theorem]{Lemma}
\newtheorem{corollary}[theorem]{Corollary}
\newtheorem{proposition}[theorem]{Proposition}
\newtheorem{conjecture}[theorem]{Conjecture}
\newtheorem{problem}[theorem]{Problem}
\newtheorem{remark}[theorem]{Remark}
\newtheorem{example}[theorem]{Example}
\newcommand{\beq}{\begin{equation}}
\newcommand{\eeq}{\end{equation}}
\newcommand{\argmax}{{\rm argmax}}
\newcommand{\MiP}{MinHOMP($H$) }
\newcommand{\MaP}{MaxHOMP($H$) }
\newcommand{\vecc}[1]{\stackrel{\leftrightarrow}{#1}}

\title{On the Complexity of the Minimum Cost Homomorphism Problem for
Reflexive Multipartite Tournaments}

\author{Gregory Gutin\thanks{Corresponding author. Department of Computer Science,
Royal Holloway University of London, Egham, Surrey TW20 OEX, UK,
gutin@cs.rhul.ac.uk and Department of Computer Science, University
of Haifa, Israel} \and Eun Jung Kim\thanks{Department of
Industrial Engineering, KAIST 373-1 Kusong-dong, Yusong-ku,
Taejon, 305-791, Republic of Korea, masquenada@kaist.ac.kr}}

\date{}

\maketitle

\begin{abstract} For digraphs $D$ and $H$, a mapping $f:\ V(D)\dom
V(H)$ is a homomorphism of $D$ to $H$ if $uv\in A(D)$ implies
$f(u)f(v)\in A(H).$ For a fixed digraph $H$, the homomorphism
problem is to decide whether an input digraph $D$ admits a
homomorphism to $H$ or not, and is denoted as HOMP($H$). Digraphs
are allowed to have loops, but not allowed to have parallel arcs.

A natural optimization version of the homomorphism problem is
defined as follows. If each vertex $u \in V(D)$ is associated with
costs $c_i(u), i \in V(H)$, then the cost of the homomorphism $f$ is
$\sum_{u\in V(D)}c_{f(u)}(u)$. For each fixed digraph $H$, we have
the {\em minimum cost homomorphism problem for} $H$ and denote it as
MinHOMP($H$). The problem is to decide, for an input graph $D$ with
costs $c_i(u),$ $u \in V(D), i\in V(H)$, whether there exists a
homomorphism of $D$ to $H$ and, if one exists, to find one of
minimum cost.

In a recent paper, we posed a problem of characterizing polynomial
time solvable and NP-hard cases of the minimum cost homomorphism
problem for acyclic multipartite tournaments with possible loops
(w.p.l.). In this paper, we solve the problem for reflexive
multipartite tournaments and demonstrate a considerate difficulty of
the problem for the whole class of multipartite tournaments w.p.l.
using, as an example, acyclic 3-partite tournaments of order 4
w.p.l.\footnote{This paper was submitted to Discrete Mathematics on
April 6, 2007}
\end{abstract}

\section{Introduction}

Our paper \cite{gutinRMS} launched research on the minimum cost
homomorphism problem for digraphs with possible loops (w.p.l.). We
characterized polynomial time solvable and NP-hard cases for some
classes of digraphs: directed cycles w.p.l., tournaments w.p.l. and
cyclic multipartite tournaments w.p.l. (a digraph is {\em cyclic} if
it contains a cycle). In \cite{gutinRMS}, we posed a problem of
characterizing polynomial time solvable and NP-hard cases of the
minimum cost homomorphism problem for two classes of digraphs
w.p.l.: semicomplete digraphs w.p.l. and acyclic multipartite
tournaments w.p.l. Such a characterization has been obtained for
semicomplete digraphs w.p.l. in our recent paper \cite{kimSJDM}. In
this paper, we characterize polynomial time solvable and NP-hard
cases of the minimum cost homomorphism problem for reflexive
multipartite tournaments and demonstrate a considerate difficulty of
the problem for the whole class of multipartite tournaments w.p.l.
using, as an example, acyclic 3-partite tournaments of order 4
w.p.l. Since the complexity of the minimum cost homomorphism problem
for undirected graphs has been completely classified
\cite{gutinEJC}, we suggest to use the bipartite representation of a
digraph to obtain results on the complexity of the minimum cost
homomorphism problem for some digraphs (see Lemma \ref{biH}).

In this paper, directed and undirected graphs may have loops, but
they do not have parallel arcs and edges. If a directed (undirected)
graph $G$ has no loops, we call $G$ {\em loopless}. If a directed
(undirected) graph $G$ has a loop at every vertex, we call $G$ {\em
reflexive}. When we wish to stress that a family of digraphs have
digraphs with loops, we will say that we deal with digraphs {\em
with possible loops (w.p.l.)} For an undirected graph $G$, $V(G)$
and $E(G)$ denote its vertex and edge sets, respectively. For a
digraph $G$, $V(G)$ and $A(G)$ denote its vertex and arc sets,
respectively.

Given directed (undirected) graphs $G$ and $H$, a {\em
homomorphism of $G$ to $H$} is a mapping $f:\ V(G)\dom V(H)$ such
that $f(u)f(v)$ is an arc (edge) of $H$ whenever $uv$ is an arc
(edge) of $G$. A homomorphism $f$ of $G$ to $H$ is also called an
{\em $H$-coloring} of $G$, and $f(x)$ is called the {\em color} of
the vertex $x$ in $G$. Let $H$ be a fixed directed or undirected
graph. The {\em homomorphism problem} for $H$, denoted as
HOMP($H$), asks whether a directed or undirected input graph $G$
admits a homomorphism to $H.$

We can strengthen HOMP($H$)  by imposing a restriction on the
image $f(u)$ of each vertex $u\in V(G)$ or by introducing costs
for the assignment of a color to a vertex $u\in V(G)$. For a fixed
directed or undirected graph $H$, the {\em list homomorphism
problem} for $H$, denoted as ListHOMP($H$), asks whether a
directed or undirected input graph $G$ with lists (sets) $L_u
\subseteq V(H), u \in V(G)$ admits a homomorphism $f$ to $H$ in
which $f(u) \in L_u$ for each $u \in V(G)$. In the {\em minimum
cost homomorphism problem}, for a fixed directed or undirected
graph $H$, we are given an input graph $G$ and associated costs
$c_i(u),i\in V(H)$ for each vertex $u\in V(G)$. The problem,
denoted as MinHOMP($H$), is to decide whether $G$ admits a
homomorphism to $H$ and, if one exists, to find one of minimum
cost. Here, the cost of a homomorphism $f$ of $G$ to $H$ is given
by $\sum_{u\in V(G)}c_{f(u)}(u)$. Note that the list homomorphism
problem is a strengthening  of the homomorphism problem, and the
minimum cost homomorphism problem is again a strengthening of the
list homomorphism problem. Thus, in particular, if MinHOMP($H$) is
polynomial time solvable, then ListHOMP($H$) and HOMP($H$) are
polynomial time solvable as well.

The minimum cost homomorphism problem was introduced in
\cite{gutinDAM154a}, where it was motivated by a real-world problem
in defence logistics. We believe it offers a practical and natural
model for optimization of weighted homomorphisms. Apart from the
list homomorphism problem, special cases of MinHOMP($H$) include the
general optimum cost chromatic partition problem, which has been
intensively studied \cite{halld2001,jansenJA34}, and has a number of
its own special cases \cite{jiangGT32,kroon1997} and applications
\cite{kroon1997,supowitCAD6}.

For undirected graphs, the complexities of the problems HOMP($H$),
ListHOMP($H$), MinHOMP($H$) for a fixed graph $H$ have been fully
classified. In \cite{hellJCT48}, Hell and Ne\v{s}et\v{r}il proved
that the problem HOMP($H$) is NP-complete if $H$ is a loopless
non-bipartite graph and it is polynomial time solvable, otherwise.
For ListHOMP($H$), Feder, Hell and Huang \cite{federJGT} proved that
the problem is polynomial time solvable if $H$ is a bi-arc graph,
and it is NP-complete, otherwise. Gutin, Hell, Rafiey and Yeo
\cite{gutinEJC} proved that MinHOMP($H$) is polynomial time solvable
if $H$ is a reflexive proper interval graph or a loopless proper
interval bigraph, and it is NP-hard, otherwise.

On the other hand, it turns out that the task of obtaining a
dichotomy classification for these problems requires much more
effort when it comes to directed graphs. Dichotomy classifications
for HOMP($H$) have been obtained for some limited special digraph
classes. For example, given a semicomplete digraph $H$, HOMP($H$) is
polynomial time solvable if $H$ contains at most one cycle, and is
NP-complete otherwise \cite{bangSIAMJDM1}. There are a few more
simple digraph classes for which polynomial solvability of the
homomorphism problem has been stated, see \cite{hell2004} for
details. Note that digraphs with at least one loop are of no
interest since we have a trivial homomorphism of any digraph to
those.

The complexity of ListHOMP($H$) for digraphs was studied only in two
papers. The existence of a dichotomy of ListHOMP(H) follows from the
main result in \cite{bulatov}, where difficulties in obtaining the
actual dichotomy are stressed. To the best of the authors'
knowledge, \cite{federMAN} is the only attempt so far to obtain a
concrete dichotomy classification of ListHOMP($H$) for digraphs. The
authors of \cite{federMAN} conjectured that for a reflexive digraph,
ListHOMP($H$) is polynomial time solvable if and only if $H$ has a
proper ordering.

For MinHOMP($H$), a complete dichotomy for a general digraph $H$ has
not been established yet. Nonetheless, for some special classes of
loopless digraphs such as semicomplete digraphs \cite{gutinDAM154b}
and semicomplete multipartite digraphs \cite{gutinDAM, gutinSIDMA},
full dichotomy classifications for MinHOMP($H$) have been obtained.
The problem MinHOMP($H$) for digraphs w.p.l. was studied in
\cite{gutinRMS,kimSJDM}, where dichotomy classifications were
obtained for some classes of digraphs including directed cycles and
semicomplete digraphs w.p.l.

\section{Additional Terminology and Notation}

Let $D$ be a digraph. The {\em converse} of $D$ is the digraph
obtained from $D$ by replacing every arc $xy$ with the arc $yx$. If
$xy \in A(D)$, we say that $x$ {\em dominates} $y$ and $y$ {\em is
dominated} by $x$, denoted by $x \rightarrow y$. For sets $X, Y
\subseteq V(D)$, $X \rightarrow Y$ means that $x \rightarrow y$ for
each $x \in X$, $y \in Y$.

For a digraph $H$, let $H[X]$ denote a subdigraph induced by $X
\subseteq V(H)$. For any pair of vertices of a directed graph $H$,
we say that $u$ and $v$ are {\em adjacent} if $u\rightarrow v$ or $v
\rightarrow u$, or both. The {\em underlying graph} $U(H)$ of a
directed graph $H$ is the undirected graph obtained from $H$ by
disregarding all orientations and deleting one edge in each pair of
parallel edges. A digraph $H$ is {\em connected} if $U(H)$ is
connected. The {\em components} of $H$ are the subdigraphs of $H$
induced by the vertices of components of $U(H)$. For a digraph
$H=(V,A)$, $BG(H)=(V_1,V_2;E)$ denotes the bipartite graph with
partite sets $V_1=\{v_1:\ v\in V\},\ V_2=\{v_2:\ v\in V\}$ such that
$u_1w_2\in E$ if and only if $uw\in A$. We call $BG(H)$ the {\em
bipartite representation} of $H$ \cite{bang2000}.

By a {\em directed path (cycle)} we mean a simple directed path
(cycle) (i.e., with no self-crossing). We assume that a directed
cycle has at least two vertices. In particular, a loop is not a
cycle. A directed cycle with $k$ vertices is called a {\em
directed $k$-cycle} and denoted by $\vec{C}_k.$ A digraph $H$ is
an {\em extension} of a digraph $D$ if $H$ can be obtained from
$D$ by substituting every vertex $u\in V(D)$ with a set $S_u$ of
independent vertices such that $u \rightarrow v$ in $D$ implies
$S_u \rightarrow S_v$ in $H$.

A loopless digraph $D$ is a {\em tournament} ({\em semicomplete
digraph}) if there is exactly one arc (at least one arc) between
every pair of vertices. We will consider {\em semicomplete
digraphs with possible loops (w.p.l.)}, i.e., digraphs obtained
from semicomplete digraphs by appending some number of loops
(possibly zero loops). A {\em $k$-partite tournament} ({\em
semicomplete $k$-partite digraph}) is a digraph obtained from a
complete $k$-partite graph by replacing every edge $xy$ with one
of the two arcs $xy,yx$ (with at least one of the arcs $xy,yx$).
It is also called a {\em multipartite tournament} (a {\em
semicomplete multipartite digraph}). An acyclic tournament on $p$
vertices is denoted by $TT_p$ and called a {\em transitive
tournament}. The vertices of a transitive tournament $TT_p$ can be
labeled $1,2,\ldots ,p$ such that $ij\in A(TT_p)$ if and only if
$1\le i<j\le p.$ By $TT^-_p$ $(p\ge 2$), we denote $TT_p$ without
the arc $1p.$ For an acyclic digraph $H$, an ordering
$u_1,u_2,\ldots ,u_p$ is called {\em acyclic} if $u_i\dom u_j$
implies $i<j.$


As usual $K_{n,m}$ denotes a complete bipartite graph with bipartite
sets of cardinalities $n$ and $m.$ By $\vec{K}_{n,m}$ we denote the
digraph obtained from $K_{n,m}$ by orienting all edges from the
bipartite set of cardinality $n$ to the bipartite set of cardinality
$m$. For a digraph $H$, the {\em reflexive closure} $RC(H)$ is the
digraph obtained from $H$ by adding a loop to every vertex of $H$
without a loop.

An undirected graph $G$ is called an {\em interval graph} if it can
be represented by a family of intervals on the real line so that
each vertex $u\in V(G)$ corresponds to an interval $I_u$, in which
two vertices $u$ and $v$ in $V(G)$ are adjacent if and only if $I_u$
and $I_v$ overlap. If the intervals can be chosen in an
inclusion-free way, we call the graph a {\em proper interval graph}.
A bipartite graph $G$ with bipartition $S\cup T$ is called an {\em
interval bigraph} if each partite set can be represented by a family
of intervals on the real line so that each vertex $u\in V(G)$
corresponds to an interval $I_u$, in which two vertices $u\in S$ and
$v\in T$ are adjacent if and only if $I_u$ and $I_v$ overlap. If
each family of intervals can be chosen to be inclusion-free, we call
the graph a {\em proper interval bigraph.}

\section{Preliminary Results}

In this section, we present some new and known results which will be
frequently used to prove either NP-hardness or polynomial time
solvability in this paper. The following lemma is an obvious basic
observation often used to obtain dichotomies. This lemma is
generally applicable even when $H$ is w.p.l.

\begin{lemma}\cite{gutinDAM154b}\label{reduction}
Let $H'$ be an induced subdigraph of a digraph $H$. If
MinHOMP($H'$) is NP-hard, then MinHOMP($H$) is also NP-hard.
\end{lemma}

\begin{remark}\label{connrem}
For a fixed directed or undirected graph $H$, let MinHOMPc($H$) be
the same problem as MinHOMP($H$), but all inputs of MinHOMPc($H$)
are connected. Notice that if MinHOMPc($H$) is NP-hard, then so is
MinHOMP($H$) as a more general problem. If MinHOMPc($H$) is
polynomial time solvable, we can solve MinHOMP($H$) in polynomial
time as well. Indeed, let $G_1,G_2,\ldots ,G_g$ be components of
an input $G$ of MinHOMP($H$). Observe that to solve MinHOMP($H$)
for $G$ it suffices to solve MinHOMPc($H$) for each $G_i$,
$i=1,2,\ldots ,g$ separately. Thus, whenever we prove NP-hardness
or polynomial time solvability of MinHOMP($H$), we may assume that
all inputs are connected.
\end{remark}

Using Lemma \ref{reduction}, we can prove NP-hardness of
MinHOMP($H$) by showing that $H$ contains as a subdigraph $H'$ for
which NP-hardness of MinHOMP($H'$) is known already. Hence, we may
concentrate on some small-sized `essential' digraphs and try to
construct a polynomial reduction from an NP-complete problem to the
minimum cost homomorphism problem with respect to these essential
digraphs. Below (see Lemma \ref{biH}) we suggest another way to
prove NP-hardness of MinHOMP($H$) in which we can easily utilize the
known complexity results of MinHOMP($H$) and, thus avoid building a
polynomial reduction from another NP-compete problem.

Let $L$ be a bipartite graph with ordered bipartite sets $I$ and
$J$. We define the following modification of MinHOMP($L$): given a
bipartite graph $G$ with ordered bipartite sets $X,Y$, check
whether there exists a homomorphism $f$ of $G$ to $L$ such that
$f(X)\subseteq I$ and $f(Y)\subseteq J$ and, if one exists, find
such a homomorphism of minimum cost. The new problem is denoted by
MinHOMPs($L$).

\begin{lemma}\label{red1}
Let $L$ be a bipartite graph such that MinHOMP($L$) is NP-hard.
Then MinHOMPs($L$) is NP-hard as well.
\end{lemma}
\pf Let $L_1,L_2,\ldots , L_{\ell}$ be components of $L$ and let
$I_p,J_p$ be ordered bipartite sets of $L_p,\ p=1,2,\ldots ,\ell$.
We can reduce MinHOMP($L$) to MinHOMPs($L$) as follows. Let a
bipartite graph $G$ with bipartite sets $X,Y$ be an input of
MinHOMP($L$). By Remark \ref{connrem}, we may assume that $G$ is
connected. For each $p\in \{1,2,\ldots ,\ell\}$, solve
MinHOMPs($L_p$) with bipartite sets of $G$ ordered as $X,Y$ and
then MinHOMPs($L_p$) with bipartite sets of $G$ ordered as $Y,X.$
Among the optimal solutions of the $2\ell=O(1)$ problems choose
the minimum cost one.\qed

\2

Note that according to the definition of the bipartite
representation $BG(H)$ of a digraph $H$, any loop at $x \in V(H)$
yields an edge $x_1x_2$ in $BG(H)$.

\begin{lemma}\label{biH}
Let $H$ be a digraph w.p.l. If MinHOMP($BG(H)$) is NP-hard, then
MinHOMP($H$) is also NP-hard.
\end{lemma}
\pf By Lemma \ref{red1}, it suffices to show that MinHOMPs($BG(H)$)
can be reduced to MinHOMP($H$). Let the ordered bipartite sets of
$BG(H)$ be $X_1=\{x_1:\ x\in V(H)\}$ and $X_2=\{x_2:\ x\in V(H)\}$
and let $G$ be an input bipartite graph with ordered bipartite sets
$S,T$. Construct a digraph $D$ by orienting all the edges in $E(G)$
from $S$ to $T$. We set the cost of homomorphism of $D$ to $H$ as
follows: $c_x(u)=c_{x_1}(u)$ if $u \in S$, $c_x(u)=c_{x_2}(u)$ if $u
\in T$.

Let $f$ be a homomorphism of $G$ to $BG(H)$ such that $f(S)\subseteq
X_1$ and $f(T)\subseteq X_2$. Then we can define a corresponding
homomorphism $f'$ of $D$ to $H$ with the same cost by setting
$f'(u)=x$ if $f(u)=x_1$ or $f(u)=x_2$ for each $u \in V(D)$. For an
arc $uv \in A(D)$, let $f'(u)=x$ and $f'(v)=y$. Then $x_1y_2$ is an
edge of $BG(H)$ since $f$ is a homomorphism of $G$ to $BG(H)$. Thus,
$xy$ is an arc of $H$ by the definition of $BG(H)$. It follows that
$f'$ is a homomorphism of $D$ to $H$. It is easy to see the cost of
$f'$ is the same as $f$.

Conversely, let $h'$ be a homomorphism of $D$ to $H$. Then we can
define a corresponding homomorphism $h$ of $G$ to $BG(H)$ (such that
$h(S)\subseteq X_1$ and $h(T)\subseteq X_2$) with the same cost by
setting $h(u)=x_1$ if $h(u)=x$ and $u \in S$, $h(u)=x_2$ if $h(u)=x$
and $u \in T$. With a similar argument, we conclude that $h$ is a
homomorphism of $G$ to $BG(H)$ with the same cost as $h'$.

It follows that if MinHOMP($BG(H)$) is NP-hard, MinHOMP($H$) is
NP-hard as well. \qed

\2

Unfortunately, the converse of Lemma \ref{biH} does not hold in
general. Indeed, let $V(H)=\{1,2\}$  and let $A(H)=\{12,21,11\}$.
Observe that MinHOMP($H$) is NP-hard (as the problem is equivalent
to the maximum independent set problem, see \cite{gutinRMS}), but
MinHOMP($BG(H)$) is polynomial time solvable (by Theorems
\ref{undich} and \ref{pibforbidden}).

The following theorem is the main result of \cite{gutinEJC}.

\begin{theorem}\label{undich}
Let $H$ be a connected graph with possible loops. If $H$ is a
reflexive proper interval graph or a loopless proper interval
bigraph, then the problem MinHOMP($H$) is polynomial time solvable.
In all other cases, the problem MinHOMP($H$) is NP-hard.
\end{theorem}

In the light of Theorem \ref{undich}, Lemma \ref{biH} is useful to
prove NP-hardness of MinHOMP($H$) for many digraphs $H$. If the
bipartite representation $BG(H)$ of a digraph $H$ is not a proper
interval bigraph, we immediately have NP-hardness of MinHOMP($H$).
To see whether a bipartite graph is a proper interval graph or not,
we have the following characterization of a proper interval bigraph
from \cite{hellJGT46}. Before stating the theorem, we give a number
of necessary definitions.

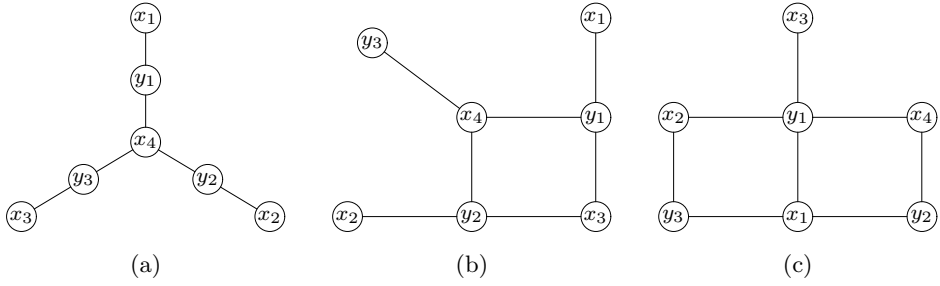
\begin{figure}
\unitlength 0.330mm \linethickness{0.4pt} \noindent
\begin{picture}(   120.00,   100.00)
\put(    60.00,    90.00){\circle{12.0}} \put(  60.000,
90.000){\makebox(0,0){{\scriptsize $x_1$}}} \put(   110.00,
10.00){\circle{12.0}} \put( 110.000,
10.000){\makebox(0,0){{\scriptsize $x_2$}}} \put(    10.00,
10.00){\circle{12.0}} \put(  10.000,
10.000){\makebox(0,0){{\scriptsize $x_3$}}} \put(    60.00,
40.00){\circle{12.0}} \put(  60.000,
40.000){\makebox(0,0){{\scriptsize $x_4$}}} \put(    60.00,
65.00){\circle{12.0}} \put(  60.000,
65.000){\makebox(0,0){{\scriptsize $y_1$}}} \put(    85.00,
25.00){\circle{12.0}} \put(  85.000,
25.000){\makebox(0,0){{\scriptsize $y_2$}}} \put(    35.00,
25.00){\circle{12.0}} \put(  35.000,
25.000){\makebox(0,0){{\scriptsize $y_3$}}} \drawline(  60.000,
84.000)(  60.000,  71.000) \drawline(  60.000,  59.000)(  60.000,
46.000) \drawline( 104.855,  13.087)(  90.145,  21.913) \drawline(
79.855,  28.087)(  65.145,  36.913) \drawline(  15.145,  13.087)(
29.855,  21.913) \drawline(  40.145,  28.087)(  54.855,  36.913)
\put(60,-10){\makebox(0,0){{\footnotesize (a)}}}
\end{picture} \hspace{-0.01cm}
\linethickness{0.4pt}
\begin{picture}(   120.00,   100.00)
\put(   110.00,    90.00){\circle{12.0}} \put( 110.000,
90.000){\makebox(0,0){{\scriptsize $x_1$}}} \put(    10.00,
10.00){\circle{12.0}} \put(  10.000,
10.000){\makebox(0,0){{\scriptsize $x_2$}}} \put(   110.00,
10.00){\circle{12.0}} \put( 110.000,
10.000){\makebox(0,0){{\scriptsize $x_3$}}} \put(    60.00,
50.00){\circle{12.0}} \put(  60.000,
50.000){\makebox(0,0){{\scriptsize $x_4$}}} \put(   110.00,
50.00){\circle{12.0}} \put( 110.000,
50.000){\makebox(0,0){{\scriptsize $y_1$}}} \put(    60.00,
10.00){\circle{12.0}} \put(  60.000,
10.000){\makebox(0,0){{\scriptsize $y_2$}}} \put(    20.00,
80.00){\circle{12.0}} \put(  20.000,
80.000){\makebox(0,0){{\scriptsize $y_3$}}} \drawline( 110.000,
84.000)( 110.000,  56.000) \drawline( 110.000,  44.000)( 110.000,
16.000) \drawline( 104.000,  10.000)(  66.000,  10.000) \drawline(
60.000,  16.000)(  60.000,  44.000) \drawline(  66.000,  50.000)(
104.000,  50.000) \drawline(  16.000,  10.000)(  54.000,  10.000)
\drawline(  24.800,  76.400)(  55.200,  53.600)
\put(60,-10){\makebox(0,0){{\footnotesize (b)}}}
\end{picture}  \hspace{-0.01cm}
\linethickness{0.4pt}
\begin{picture}(   120.00,   100.00)
\put(    60.00,    10.00){\circle{12.0}} \put(  60.000,
10.000){\makebox(0,0){{\scriptsize $x_1$}}} \put(    10.00,
50.00){\circle{12.0}} \put(  10.000,
50.000){\makebox(0,0){{\scriptsize $x_2$}}} \put(    60.00,
90.00){\circle{12.0}} \put(  60.000,
90.000){\makebox(0,0){{\scriptsize $x_3$}}} \put(   110.00,
50.00){\circle{12.0}} \put( 110.000,
50.000){\makebox(0,0){{\scriptsize $x_4$}}} \put(    60.00,
50.00){\circle{12.0}} \put(  60.000,
50.000){\makebox(0,0){{\scriptsize $y_1$}}} \put(   110.00,
10.00){\circle{12.0}} \put( 110.000,
10.000){\makebox(0,0){{\scriptsize $y_2$}}} \put(    10.00,
10.00){\circle{12.0}} \put(  10.000,
10.000){\makebox(0,0){{\scriptsize $y_3$}}} \drawline(  60.000,
16.000)(  60.000,  44.000) \drawline(  60.000,  56.000)(  60.000,
84.000) \drawline(  16.000,  50.000)(  54.000,  50.000) \drawline(
10.000,  44.000)(  10.000,  16.000) \drawline( 104.000,  50.000)(
66.000,  50.000) \drawline( 110.000,  44.000)( 110.000,  16.000)
\drawline(  54.000,  10.000)(  16.000,  10.000) \drawline( 66.000,
10.000)( 104.000,  10.000)
\put(60,-10){\makebox(0,0){{\footnotesize (c)}}}
\end{picture}

\mbox{ }

\caption{A bipartite claw  (a), a bipartite net (b) and a bipartite
tent (c).}\label{fig}
\end{figure}

A bipartite graph $H$ with vertices $x_1,x_2,x_3,x_4,y_1,y_2,y_3$
is called

{\em a bipartite claw } if its edge set
$E(H)=\{x_4y_1,y_1x_1,x_4y_2,y_2x_2, x_4y_3, y_3x_3\};$

{\em a bipartite net} if its edge set
$E(H)=\{x_1y_1,y_1x_3,y_1x_4,x_3y_2,x_4y_2,y_2x_2,y_3x_4\};$

{\em a bipartite tent } if its edge set
$E(H)=\{x_1y_1,y_1x_3,y_1x_4,x_3y_2,x_4y_2,y_2x_2,y_3x_4\}.$

See Figure \ref{fig}.

\begin{theorem}\label{pibforbidden}
A bipartite graph $H$ is a proper interval bigraph if and only if
it does not contain an induced cycle of length at least six, or a
bipartite claw, or a bipartite net, or a bipartite tent.
\end{theorem}

Above we considered some approaches to prove MinHOMP($H$) is NP-hard
for a given digraph $H$. The theorem given below is very useful to
prove that MinHOMP($H$) is polynomial time solvable for some
digraphs $H$.

Let $1,2,\ldots, p$ be an ordering of the vertices of a digraph
$H$ and let $e=ik$ and $f=js$ be a pair of arcs in $H.$ The {\em
minimum} ({\em maximum}) of $e$ and $f$ is
$\min\{e,f\}=\min\{i,j\}\min\{k,s\}$
($\max\{e,f\}=\max\{i,j\}\max\{k,s\}$). Notice that $\min\{e,f\}$
and $\max\{e,f\}$ are not necessarily arcs in $H.$ A pair $e,f$ is
{\em non-trivial} if $\{\min\{e,f\},\max\{e,f\}\}\neq \{e,f\}.$ An
ordering $1,2,\ldots, p$ of $V(H)$ is a {\em Min-Max ordering} if
both $\min\{e,f\}$ and $\max\{e,f\}$ are arcs in $H$ for each
non-trivial pair $e,f$ of arcs in $H.$

\begin{theorem}\label{mmth}\cite{gutinDAM154b,gutinRMS}
Let $H$ be a digraph. If $V(H)$ has a Min-Max ordering, then
MinHOMP($H$) is polynomial time solvable.
\end{theorem}

We close this section by providing some more lemmas relevant to
proving polynomial solvability.

\begin{lemma}\label{cyclepol}\cite{gutinDAM154b}
For $H=\vec{C}_k$, $k\ge 2$, MinHOMP($H$) is polynomial time
solvable.
\end{lemma}

\begin{lemma}\label{lemext}\cite{gutinDAM}
Let $H$ be a loopless digraph. If MinHOMP($H$) is polynomial time
solvable then, for each extension $H'$ of $H$, MinHOMP($H'$) is also
polynomial time solvable.
\end{lemma}

\section{Reflexive Multipartite Tournaments}

In this section, we present a dichotomy classification of \MiP
when $H$ is a reflexive multipartite tournament. The following
theorem is the main result of this section.

\begin{theorem}\label{rkdich}
Let $H$ be a reflexive $k$-partite tournament, $k\geq 2$. If $H$ is
$RC(TT_k)$, $RC(TT_{k+1}^-)$, $RC(\vec{K}_{1,2})$ or
$RC(\vec{K}_{2,1})$, then MinHOMP($H$) is polynomial time solvable.
Otherwise, MinHOMP($H$) is NP-hard.
\end{theorem}

We start by stating a lemma proved in \cite{gutinRMS}. It implies
that whenever we have a reflexive cycle in $H$, \MiP is NP-hard.

\begin{lemma}\label{3cycle}
Let  $H$ be a digraph obtained from $\vec{C}_k$, $k\ge 3$, by adding
at least one loop. Then MinHOMP($H$) is NP-hard.
\end{lemma}

The following theorem was also proved in \cite{gutinRMS}.

\begin{theorem}\label{tdich}
Let $T$ be a tournament w.p.l. If $H$ is an acyclic tournament
w.p.l. or $H=\vec{C}_3$, then MinHOMP($H$) is polynomial time
solvable. Otherwise, MinHOMP($H$) is NP-hard.
\end{theorem}

By the two lemmas given below, \MiP is NP-hard if $H$ has a
partite set consisting of three or more vertices.

\begin{lemma}\label{3loops}
Let $H'$ be a digraph with $V(H')=\{u,v,w,z\}$ and
$A(H')=\{zu,zv,zw\}\cup B$. If $\{uu,vv,ww\}\subseteq B \subseteq
\{uu,vv,ww,zz\}$ and $H$ is $H'$ or its converse, \MiP is NP-hard.
Otherwise, it is polynomial time solvable.
\end{lemma}
\pf Suppose $B$ satisfies the condition of the statement. Consider
the bipartite graph $BG(H)$. The subgraph induced by
$\{u_1,u_2,v_1,v_2,w_1,w_2\}$ together with $z_1$ (when $z$
dominates $u,v,w$) or $z_2$ (when $z$ is dominated by $u,v,w$) is a
bipartite claw. Then, $BG(H)$ is not a proper interval bigraph by
Theorem \ref{pibforbidden}, which implies that MinHOMP($BG(H)$) is
NP-hard by Theorem \ref{undich}. Thus by Lemma \ref{biH},
MinHOMP($H$) is NP-hard. For all the other cases, we can easily
check that $H$ has a Min-Max ordering. \qed

\begin{lemma}\label{3loops2}
Let $H'$ be a reflexive digraph with $V(H')=\{u,v,w,z\}$ and
$A(H')=\{zu,zv,wz,zz,uu,vv\}$ and let $H''=RC(H')$. Let $H$ be
$H'$ or its converse, or $H''$ or its converse. Then MinHOMP($H$)
is NP-hard.
\end{lemma}
\pf Consider the bipartite graph $BG(H)$. The subgraph induced by
$\{w_1,z_1,z_2,u_1,u_2,v_1,v_2\}$ is a bipartite claw. Then, by
Theorems  \ref{pibforbidden} and \ref{undich}, MinHOMP($BG(H)$) is
NP-hard. By Lemma \ref{biH}, MinHOMP($H$) is NP-hard again. \qed

\2

 We have one more structure which leads to NP-hardness of \MiP.

\begin{lemma}\label{ac1234}
Let $H'$ be given by $V(H')=\{u,w,v,z\},\
A(H')=\{uw,wv,vz,uz,wz\}\cup B$, where $B$ is $\{uu,ww,vv,zz\}$.
Let $H$ be $H'$ or its converse. Then \MiP is NP-hard.
\end{lemma}
\pf Consider the bipartite graph $BG(H)$. Observe that the subgraph
induced by $V(BG(H))\setminus \{u_2\}$ is a bipartite tent. \qed

\2

On the other hand, the following lemma describes when \MiP is
polynomial time solvable.

\begin{lemma}\label{polyrsmd}
If $H$ is $RC(TT_p)$ ($p\ge 1$), $RC(TT_p^-)$ $(p\ge 3)$,
$RC(\vec{K}_{1,2})$ or $RC(\vec{K}_{2,1})$, then MinHOMP($H$) is
polynomial time solvable.
\end{lemma}
\pf The first case is trivial. To show the case $H=RC(TT^-_p)$,
label the vertices of $TT^-_p$ by $1,2,\ldots p$ such that $ij\in
A(TT^-_p)$ if and only if $1\le i<j\le p$, but $ij\neq 1p.$ Observe
that $1,2,\ldots ,p$ is a Min-Max ordering since $1p$ can be neither
the minimum nor the maximum of a non-trivial pair of arcs. Let
$H=RC(\vec{K}_{1,2})$, $V(H)=\{1,2,3\}$ and
$A(H)=\{12,13,11,22,33\}.$ Then $2,1,3$ is a Min-Max ordering since
there is no pair of non-trivial arcs. The case $H=RC(\vec{K}_{2,1})$
is similar. \qed

\2

Now we are ready to prove Theorem \ref{rkdich}.

{\bf Proof of Theorem \ref{rkdich}:} Let $S_1,\ldots ,S_k$ be the
partite sets of $H$. Assume first that there are at least two
partite sets with at least 2 vertices each. Without loss of
generality, let both $S_1$ and $S_2$ have two or more vertices.

(a) Assume first that there is a vertex $u\in S_1$ such that
$N^+(u) \cap S_2 \neq \emptyset$ and $N^-(u) \cap S_2 \neq
\emptyset$. There are the following three cases to consider.

{\bf Case 1:} There is a vertex $v \in V(S_1)$ such that $N^+(u)
\cap N^-(v) \cap S_2 \neq \emptyset$ and $N^-(u) \cap N^+(v) \cap
S_2 \neq \emptyset$. Choose two vertices $w$ and $z$ from $N^+(u)
\cap N^-(v) \cap S_2$ and $N^-(u) \cap N^+(v) \cap S_2$,
respectively. Then $u,v,w,$ and $z$ form $\vec{C}_4$, implying
that MinHOMP($H$) is NP-hard by Lemmas \ref{3cycle} and
\ref{reduction}.

{\bf Case 2:} There is a vertex $v \in S_1$ such that exactly one
of the two sets $N^+(u) \cap N^-(v) \cap S_2$ and $N^-(u) \cap
N^+(v) \cap S_2$ is nonempty. Without loss of generality, we may
assume that $N^+(u) \cap N^-(v) \cap S_2 \neq \emptyset$ and
$N^-(u) \cap N^+(v) \cap S_2 = \emptyset$. Choose a vertex $w \in
N^+(u) \cap N^-(v) \cap S_2$ and a vertex $z \in N^-(u)\cap S_2$.
Note that $z$ also dominates $v$.

Let $H_s$ be the digraph induced by $u,v,w$ and $z$ and recall that
$BG(H_s)$ denotes the bipartite representation of $H_s$. Then
$z_1,u_1,u_2,w_1,w_2$ and $v_2$ induce a cycle of length six, thus
$BG(H_s)$ is not a proper interval bigraph by Theorem
\ref{pibforbidden}. By Lemmas \ref{biH} and \ref{reduction},
MinHOMP($H$) is NP-hard.

{\bf Case 3:} For every vertex $v \in S_1$, we have $N^+(v)\cap
S_2=N^+(u)\cap S_2$ and $N^-(v)\cap S_2 = N^-(u)\cap S_2$. Choose
two vertices $w$ and $z$ from $N^+(u) \cap S_2$ and $N^-(u) \cap
S_2$, respectively and a vertex $v\in S_1\setminus \{u\}$. Let $H_s$
be the digraph induced by $u,v,w$ and $z$. Then
$z_1,u_1,u_2,v_1,v_2$ and $w_2$ induce a cycle of length six, and,
thus, $BG(H_s)$ is not a proper interval bigraph by Theorem
\ref{pibforbidden}. By Lemmas \ref{biH} and \ref{reduction},
MinHOMP($H$) is NP-hard.

(b) Now assume that $S_1$ dominates $S_2$. Choose $u,v \in S_1$ and
$w,z\in S_2$. Let $H_s$ be the digraph induced by $u,v,w$ and $z$.
Then $u_1,u_2,v_1,v_2,w_1,w_2$ and $z_2$ induce a bipartite net and,
thus, $BG(H_s)$ is not a proper interval bigraph by Theorem
\ref{pibforbidden}. By Lemmas \ref{biH} and \ref{reduction},
MinHOMP($H$) is NP-hard.

By (a) and (b), if there are two or more partite sets of
cardinality larger than 1, MinHOMP($H$) is NP-hard. Furthermore,
MinHOMP($H$) is also NP-hard when any partite set has three or
more vertices, by Lemmas \ref{3loops} and \ref{3loops2}. Hence, we
assume that there is at most one partite set of cardinality 2 and
all the other partite sets consist of a single vertex.

When there is no partite set of cardinality 2, then $H$ is a
reflexive tournament. In this case, we have a dichotomy
classification by Theorem \ref{tdich} from \cite{gutinRMS}, which
asserts that MinHOMP($H$) is polynomial time solvable if $H$ is a
reflexive acyclic tournament, and it is NP-hard otherwise.

Consider the case when there is a unique partite set $S_i$ of
cardinality 2. Let $u,v$ be the two vertices of $S_i$. If $H$
contains a cycle as an induced subdigraph, MinHOMP($H$) is NP-hard
by Lemmas \ref{3cycle} and \ref{reduction}. Hence, let us assume
that $H$ is acyclic. Then there are the following three cases to
consider.

{\bf Case 1:} There are two vertices $w$ and $z$, each from a
distinct partite set, such that $w$ dominates $S_i$ and $z$ is
dominated be $S_i$. Note that $w$ dominates $z$ since we assumed
that $H$ is acyclic. Let $H_s$ be the digraph induced by $u,v,w$.
Then $w_1,w_2,u_1,u_2,v_1,v_2$ and $z_2$ induce a bipartite tent,
thus $BG(H_s)$ is not a proper interval bigraph by Theorem
\ref{pibforbidden}. By Lemmas \ref{biH} and \ref{reduction},
MinHOMP($H$) is NP-hard.

{\bf Case 2:} The partite set $S_i$ either dominates all the other
partite sets or is dominated by all the other partite sets. If
$k=2$, then we arrive to two polynomial cases by Lemma
\ref{polyrsmd}. Thus, we may assume that $k \geq 3$. Without loss of
generality, we may assume that $S_i$ is dominated by all the other
partite sets. Let $w,z$ be two vertices from partite sets other than
$S_i$. We may assume that $w$ dominates $z$. Let $H_s$ be the
digraph induced by $u,v,w$ and $z$. Then $w_1,w_2,u_1,u_2,v_1,v_2$
and $z_1$ induce a bipartite net, thus $BG(H_s)$ is not a proper
interval bigraph by Theorem \ref{pibforbidden}. By Lemmas \ref{biH}
and \ref{reduction}, MinHOMP($H$) is NP-hard.

{\bf Case 3:} There is a vertex $w$ from a partite set other than
$S_i$ such that $u$ dominates $w$ and $w$ dominates $v$. If there
is another vertex $z$ which either dominates or is dominated by
both $u$ and $v$, we respectively have $z\dom w$ or $w\dom z$
since $H$ is acyclic. Then MinHOMP($H$) is NP-hard by Lemmas
\ref{ac1234} and \ref{reduction}. So, we may assume that such a
vertex $z$ does not exist. Then $u$ dominates all the vertices $y
\in V(H)-\{u,v\}$ and $v$ is dominated by all the vertices $y \in
V(H)-\{u,v\}$ since $H$ is acyclic. Moreover, since $H$ is
acyclic, $H$ is $TT_{k+1}^-$. By Lemmas \ref{polyrsmd} and
\ref{mmth}, MinHOMP($H$) is polynomial time solvable. \qed

\section{Concluding Remarks}

In this paper, we suggested a dichotomy classification of \MiP for
reflexive multipartite tournaments. Moreover, we suggested to use
the bipartite representations of digraphs to prove \MiP is NP-hard
for some digraphs $H$.

The general case of acyclic multipartite tournaments w.p.l. remains
elusive. We suspect that the main reason for the difficulty of \MiP
for the general case is the fact that \MiP polynomial time solvable
for a large number of acyclic multipartite tournaments w.p.l. and
determining all such digraphs is not easy. The following theorem,
which we give without a proof as the proof we have is rather long
and technical, indicates variety of polynomial cases.

\begin{theorem}\label{th1}
Let $H$ be given by $V(H)=\{1,2,3,4\},\ A(H)=\{12,23,34,14,24\}\cup
B$, where $B\subseteq \{11,22,33,44\}$. If $\{33\}\subseteq
B\subseteq \{11,22,33\}$, then MinHOMP($H$) is polynomial time
solvable. Otherwise, MinHOMP($H$) is NP-hard.
\end{theorem}

Let $H'$ have vertex set $\{1,2,3,4\}$ and arc set $\{12,23,24\}\cup
B$, where $B\subseteq \{11,22,33,44\}$. We have been unable to
obtain an analog of Theorem \ref{th1} for $H'.$ For example, it is
unclear what is the complexity of the case $B=\{33,44\}.$

\end{document}